\documentclass[showpacs,floatfix,reprint,prm,superscriptaddress]{revtex4-1}
\usepackage{xfrac}
\usepackage{amssymb}
\usepackage{braket}
\usepackage{graphicx}
\usepackage{verbatim}
\usepackage{etoolbox}
\usepackage{amsfonts}
\usepackage{amsmath}
\usepackage[utf8]{inputenc}
\usepackage{mathtools}
\usepackage{soul,xcolor,colortbl}
\usepackage{hyperref} \hypersetup{colorlinks=true,citecolor=blue,linkcolor=blue,urlcolor=blue}
\makeatletter \renewcommand\frontmatter@abstractwidth{\dimexpr\textwidth\relax} \makeatother

\usepackage{multirow}
\usepackage{longtable}
\usepackage{cellspace}
\usepackage{resizegather}
\usepackage{dcolumn}
\usepackage[normalem]{ulem}

\usepackage{bm}
\usepackage{array}
\newcolumntype{C}[1]{>{\centering\arraybackslash}p{#1}}\usepackage{soul}
\definecolor{Gray}{gray}{0.85}

\usepackage{color}
\usepackage{morefloats}

\definecolor{Gray}{gray}{0.9}
\definecolor{LightCyan}{rgb}{0.88,1,1}
\definecolor{green}{rgb}{0.5451,0.2706,0.0745}

\def\DFT{{\small  DFT}}
\def\CCE{{\small  CCE}}
\def\LDA{{\small  LDA}}
\def\GGA{{\small  GGA}}
\def\PBE{{\small  PBE}}
\def\SCAN{{\small  SCAN}}

\def\MAE{{\small  MAE}}

\def\NJ{{\small  NJ}}

\def\AFLOW{{\small AFLOW}}
\def\VASP{{\small VASP}}
\def\POSCAR{{\small POSCAR}}

\def\citePROTOS{\cite{curtarolo:art121,curtarolo:art145}}
\def\tablecolumnspacing{2.3}

\begin{document}

\title{Automated coordination corrected enthalpies with AFLOW-CCE}

\author{Rico Friedrich}
\affiliation{Department of Mechanical Engineering and Materials Science, Duke University, Durham, North Carolina 27708, USA}
\affiliation{Center for Autonomous Materials Design, Duke University, Durham, North Carolina 27708, USA}
\affiliation{Institute of Ion Beam Physics and Materials Research, Helmholtz-Zentrum Dresden-Rossendorf, 01328 Dresden, Germany}
\author{Marco Esters}
\affiliation{Department of Mechanical Engineering and Materials Science, Duke University, Durham, North Carolina 27708, USA}
\affiliation{Center for Autonomous Materials Design, Duke University, Durham, North Carolina 27708, USA}
\author{Corey Oses}
\affiliation{Department of Mechanical Engineering and Materials Science, Duke University, Durham, North Carolina 27708, USA}
\affiliation{Center for Autonomous Materials Design, Duke University, Durham, North Carolina 27708, USA}
\author{Stuart Ki}
\affiliation{Department of Mechanical Engineering and Materials Science, Duke University, Durham, North Carolina 27708, USA}
\affiliation{Center for Autonomous Materials Design, Duke University, Durham, North Carolina 27708, USA}
\author{Maxwell J. Brenner}
\affiliation{Department of Mechanical Engineering and Materials Science, Duke University, Durham, North Carolina 27708, USA}
\affiliation{Center for Autonomous Materials Design, Duke University, Durham, North Carolina 27708, USA}
\author{David Hicks}
\affiliation{Department of Mechanical Engineering and Materials Science, Duke University, Durham, North Carolina 27708, USA}
\affiliation{Center for Autonomous Materials Design, Duke University, Durham, North Carolina 27708, USA}
\author{Michael J. Mehl}
\affiliation{Department of Mechanical Engineering and Materials Science, Duke University, Durham, North Carolina 27708, USA}
\affiliation{Center for Autonomous Materials Design, Duke University, Durham, North Carolina 27708, USA}
\author{Cormac Toher}
\affiliation{Department of Mechanical Engineering and Materials Science, Duke University, Durham, North Carolina 27708, USA}
\affiliation{Center for Autonomous Materials Design, Duke University, Durham, North Carolina 27708, USA}
\author{Stefano Curtarolo}
\affiliation{Center for Autonomous Materials Design, Duke University, Durham, North Carolina 27708, USA}
\affiliation{Materials Science, Electrical Engineering, Physics and Chemistry, Duke University, Durham NC, 27708, USA}
\email[]{stefano@duke.edu}

\date{\today}

\begin{abstract}

\noindent
The computational design of {materials with ionic bonds} poses a critical challenge to thermodynamic modeling since density functional theory yields inaccurate predictions of their formation enthalpies.
Progress requires leveraging physically insightful correction methods.
The recently introduced coordination corrected enthalpies (\CCE) method delivers accurate formation enthalpies with mean absolute errors close to room temperature thermal energy, \emph{i.e.} $\approx$ 25~meV/atom.
The \CCE\ scheme, depending on the number of cation-anion bonds and oxidation state of the cation, requires an automated analysis of the system to determine and apply the correction.
Here, we present \AFLOW-\CCE\ --- our implementation of \CCE\ into the \AFLOW\ framework for computational materials design.
It features a command line tool, a web interface and a Python environment.
The workflow includes a structural analysis, automatically determines oxidation numbers, and accounts for temperature effects by parametrizing vibrational contributions to the formation enthalpy per bond.

\vspace{0.2cm}
\noindent
Keywords: computational materials science, formation enthalpy, high-throughput computing

\end{abstract}

\maketitle

\section{Introduction} \label{intro}

\noindent
Materials design of {systems with ionic bonding contributions}, \emph{i.e.} compounds including elements of significantly different electronegativity, necessitates an accurate modelling of their thermodynamic stability \cite{Friedrich_CCE_2019,Lany_FERE_2008,Jain_GGAU_PRB_2011,Stevanovic_FERE_2012,curtarolo:art80_etal}.
The appropriate descriptor is the formation enthalpy \cite{monsterPGM}
--- the enthalpy difference between the compound and its elemental
references, or its recursive factorization to study multicomponent
systems \cite{unavoidable_disorder}.
For metals, high-throughput compatible \mbox{(semi-)local} \underline{d}ensity \underline{f}unctional \underline{t}heory (\DFT) is known to provide accurate results with errors significantly smaller than the thermal energy at room temperature ($\approx$ 25~meV/atom) \cite{Wolverton_first-principles_2006,curtarolo:art53}.
This fueled the construction of large materials databases
with millions of entries \cite{curtarolo:art75,curtarolo:art142,materialsproject.org,oqmd.org,Kirklin_NPJCM_2015,nomadMRS,ase,cmr_repository,Pizzi_AiiDA_2016}.
On the contrary, {ionic} materials pose a much more fundamental challenge for computational approaches. \par

As outlined in Ref.~\onlinecite{Friedrich_CCE_2019}, the formation enthalpy can be subdivided into a total energy difference between the compound and the elements plus a (small) vibrational contribution due to zero-point and thermal effects.
As long as all phases involved are chemically similar (in terms of their electronic delocalization character), standard \mbox{(semi-)}local \DFT 's systematic error cancellation allows for a good approximation of the total energy difference \cite{Wolverton_first-principles_2006,curtarolo:art53}.
This breaks down for {ionic} systems, such as oxides and nitrides \cite{Wang_Ceder_GGAU_PRB_2006,Jain_GGAU_PRB_2011,Lany_FERE_2008,Stevanovic_FERE_2012,Friedrich_CCE_2019}: little error cancellation can be expected between an ionic compound and its metallic/diatomic-gaseous references.
Consequently, computing reliable formation enthalpies \emph{ab initio} would require accurate total energies for all systems involved.
This is generally not possible within a \mbox{(semi-)local} approximation.\par

Significant efforts have been undertaken to investigate the accuracy that can be obtained from a specific
level of theory.
Many studies demonstrate that compared to experimental formation enthalpies \cite{Kubaschewski_MTC_1993,Chase_NIST_JANAF_thermochem_tables_1998,Barin_1995,Wagman_NBS_thermodyn_tables_1982}, the typical \underline{m}ean \underline{a}bsolute \underline{e}rror (\MAE) for standard functionals --- such as \LDA\ \cite{DFT,von_Barth_JPCSS_LSDA_1972} or \PBE\ \cite{PBE} --- is on the order of several hundred meV/atom \cite{Wang_Ceder_GGAU_PRB_2006,Lany_FERE_2008,Jain_GGAU_PRB_2011,Stevanovic_FERE_2012,Zhang_NPJCM_2018,Isaacs_PRM_2018,Friedrich_CCE_2019}.
For meta-generalized-gradient approximations, such as the Bayesian error estimation (mBEEF) \cite{Wellendorff_mBEEF_2014} or the \underline{s}trongly \underline{c}onstrained and \underline{a}ppropriately \underline{n}ormed (\SCAN) \cite{Perdew_SCAN_PRL_2015} functionals, an \MAE\ of about 100~meV/atom is obtained \cite{Pandey_PRB_2015,Zhang_NPJCM_2018,Isaacs_PRM_2018,Friedrich_CCE_2019}.
While computationally more demanding, hybrid functionals yield only
modest improvements over \PBE\ for transition metal oxides and
sulfides \cite{Yan_calculated_PRB_2013,Shang_PRM_2019}.
Non-self-consistent \underline{e}xact \underline{ex}change plus \underline{r}andom \underline{p}hase \underline{a}pproximation ({\small EXX}+{\small RPA}) and \underline{r}enormalized \underline{a}diabatic \underline{\PBE}\ ({\small rAPBE}) calculations on \PBE\ orbitals for small sets of about 20 oxides achieved \MAE s down to 74-95~meV/atom \cite{Yan_formation_PRB_2013,Yan_calculated_PRB_2013,Jauho_PRB_2015,Olsen_NPJCM_2019}.
In conclusion, even for the most expensive {DFT-based} approaches, no satisfactory accuracy ($\approx$ 25~meV/atom) is achieved{. Preliminary tests for MgH$_2$ indicate that Quantum Monte Carlo can achieve accurate results with an error of $\approx$20~meV/atom \cite{Pozzo_PRB_2008,Mao_QMC_2011}, although it remains to be determined whether this applies generally for all materials.}

Physically motivated empirical correction schemes parametrizing \mbox{(semi-)local} \DFT\ errors with respect to measured values are the only feasible option, achieving accurate formation enthalpies and enabling high-throughput materials design of {ionic} systems.
Initially, a correction for the oxygen reference energy of 1.36~eV per O$_2$ for \PBE\ was introduced \cite{Wang_Ceder_GGAU_PRB_2006}.
This scheme was extended to other gases such as H$_2$, N$_2$, F$_2$, and Cl$_2$, as well as sulfides for several functionals \cite{Grindy_PRB_2013,Yu_PRB_2015}.
On top of this, for systems with transition metal ions, an approach for mixing \GGA\ and \GGA+$U$ calculations was developed, reducing the \MAE\ to 45~meV/atom for a test set of 49 ternary oxides \cite{Jain_GGAU_PRB_2011}.
Leveraging this method, extensive further parameterization within a local-environment dependent approach
was found to lower the \MAE\ to 19~meV/atom \cite{Wolverton_DFTUenthalpies_prb_2014}.
A drawback is, however, that non-transition metals remain uncorrected, which can be particularly problematic for heavy $p$-block elements \cite{Friedrich_CCE_2019}.
As a complementary approach, the \underline{f}itted \underline{e}lemental-phase \underline{r}eference \underline{e}nergies (FERE) method introduces energy shifts for the elements on an equal footing to minimize the error between measured and calculated results for a large set of binary compounds \cite{Lany_FERE_2008,Stevanovic_FERE_2012}.
FERE values for many elements were calculated, yielding an MAE of 48 meV/atom when applied to a test set of 55 ternary compounds.
{Recently, correction schemes have also been extended to finite temperatures and the Gibbs free energies of solids \cite{Bartel_NCOM_2018}.}
{It should be noted that the accuracy of schemes fitted to measured values is limited by the experimental error.
In the supplementary information of our previous work \cite{Friedrich_CCE_2019}, we investigated the deviation between measured values of different collections for a large set of oxides, indicating that the typical experimental error bar is on the order of 10-20~meV/atom.
}

While the above correction methods were a major step forward for materials design, their accuracy is limited and the relative stability of polymorphs --- sometimes erroneously predicted by \DFT\ \cite{Zhang_NPJCM_2018} --- cannot be corrected.
Moreover, correction methods based on only composition can lead to incorrect thermodynamic behavior when considering activity \emph{vs.} concentration \cite{Friedrich_CCE_2019}.
To address these shortcomings, we have recently introduced a new universal method: \underline{c}oordination \underline{c}orrected \underline{e}nthalpies (\CCE) \cite{Friedrich_CCE_2019}.
This advanced correction scheme is the first to leverage structural information by assigning corrections per cation-anion bond, as well as considering the cation oxidation state.
\CCE\ achieves an \MAE\ of 27 (24)~meV/atom for a test set of 71 (7) ternary oxides (halides), on par with room temperature thermal energy ($\approx$ 25~meV/atom) \cite{Friedrich_CCE_2019}.
It can also correct the relative stability at fixed composition and avoids incorrect thermodynamic behavior by construction.

Here, we present our automated implementation of \CCE\ into the \AFLOW\ framework for computational materials design.
It identifies the number of cation-anion bonds, automatically determines oxidation numbers, and includes thermal effects by applying different corrections for designated temperatures.
\AFLOW-\CCE\ includes a command line interface, a web application, and a Python environment providing useful tools for the scientific community to automatically calculate the \CCE\ correction and formation enthalpies for a given input structure.
The article is organized as follows:
after introducing the computational details of the method, a short overview on the \CCE\ functionality
in \AFLOW\ is given.
Then, the specific analyses within the implementation are described including structural analysis, automatic determination of oxidation numbers, and the inclusion of temperature effects.
Available options for the command line interface, \CCE\ corrections
for 0~K, and \CCE@exp corrections for room temperature are discussed in details.

{\section{Computational Details} \label{comp_det}}

\noindent
The \emph{ab-initio} calculations for the exchange-correlation
functionals \LDA\ \cite{DFT,von_Barth_JPCSS_LSDA_1972}, \PBE\
\cite{PBE} and \SCAN\ \cite{Perdew_SCAN_PRL_2015} are performed with
\AFLOW\
\cite{curtarolo:art53,curtarolo:art57,curtarolo:art63,aflowPAPER,curtarolo:art110,aflowPI}
and the \underline{V}ienna
\emph{\underline{A}b-initio} \underline{S}i\-mu\-lation
\underline{P}ackage (\VASP) \cite{vasp} with settings according to Refs.~\onlinecite{Friedrich_CCE_2019,curtarolo:art104}.
Thermal contributions to the formation enthalpy are calculated using
the quasi-harmonic Debye model implemented via the
\AFLOW-\underline{A}utomatic \underline{G}ibbs \underline{L}ibrary
(AGL)
\cite{BlancoGIBBS2004,curtarolo:art96,curtarolo:art115}.

Using binary compounds $A_{x_1}Y_{x_2}$ as the fit set, the \CCE\ corrections $\delta H^{T,A^{+\alpha}}_{A-Y}$ per cation-anion $A-Y$ bond and cation oxidation state $+\alpha$ are obtained from the difference between (zero-temperature and zero-pressure) \DFT\ formation enthalpies and experimental standard formation enthalpies at temperature $T$ \cite{Friedrich_CCE_2019}:

\begin{equation}\label{fit_CCE}
\Delta_{\mathrm{f}} E^{0,\mathrm{DFT}}_{A_{x_1}Y_{x_2}}-\Delta_{\mathrm{f}} H^{\circ,T,\mathrm{exp}}_{A_{x_1}Y_{x_2}} =x_1N_{A-Y}\delta H^{T,A^{+\alpha}}_{A-Y},
\end{equation}

\noindent
where $N_{A-Y}$ is the number of nearest neighbor $A-Y$ bonds and $x_{i}$ are stoichiometries for the $i$-species.
Standard conditions are indicated by the ``$\circ$'' superscript.
$T$ can be 298.15 or 0~K, \emph{i.e.} temperature effects are included in the corrections.
A detailed justification of this is presented later.

The corrections can then be applied to any multinary compound $A_{x_1}B_{x_2}\dots Y_{x_n}$ to obtain the \CCE\  formation enthalpy $\Delta_{\mathrm{f}} H^{\circ,T,\mathrm{CCE}}_{A_{x_1}B_{x_2}\mathrm{\dots}Y_{x_n}}$:

\begin{equation}\label{apply_CCE}
\Delta_{\mathrm{f}} H^{\circ,T,\mathrm{CCE}}_{A_{x_1}B_{x_2}\mathrm{\dots}Y_{x_n}} =\Delta_{\mathrm{f}} E^{0,\mathrm{DFT}}_{A_{x_1}B_{x_2}\mathrm{\dots}Y_{x_n}} - \sum_{i=1}^{n-1} x_iN_{i-Y}\delta H^{T,i^{+\alpha}}_{i-Y},
\end{equation}

\noindent
where $N_{i-Y}$ is the number of nearest neighbor bonds between the cation $i$ and anion $Y$-species.
For multi-anion compounds \cite{Kageyama_NCOM_2018}, the corrections are summed for all anions separately in Eq.~(\ref{apply_CCE}).\par

For the \AFLOW-{\small ICSD} database, the \CCE\ methodology is applied equivalently with the compound energies partly calculated within \DFT$+U$ \cite{curtarolo:art104}.
In addition, composition dependent energy shifts are applied for the elements for which a $U$ is used to align the related reference energies with the ones calculated from \DFT$+U$.\par

The room temperature ($T_{\mathrm{r}}=298.15$~K) formation enthalpy CCE@exp \cite{Friedrich_CCE_2019} calculated from experimental formation enthalpies per bond $\delta H^{T_{\mathrm{r}},i^{+\alpha}}_{i-Y,\mathrm{exp}}$ is given by:

\begin{equation}\label{CCE_exp}
\Delta_{\mathrm{f}} H^{\circ,T_{\mathrm{r}},\mathrm{CCE@exp}}_{A_{x_1}B_{x_2}\mathrm{\dots}Y_{x_n}} = \sum_{i=1}^{n-1} x_iN_{i-Y}\delta H^{T_{\mathrm{r}},i^{+\alpha}}_{i-Y,\mathrm{exp}}.
\end{equation}

\noindent
These values provide a rough guess with an estimated average accuracy of about 250~meV/atom as obtained from a test for ternary oxides \cite{Friedrich_CCE_2019}.

\vspace{0.7cm}

{\section{Results} \label{results_discussion}}

\begin{figure*}[p]
	\centering
	\includegraphics[width=0.99\textwidth]{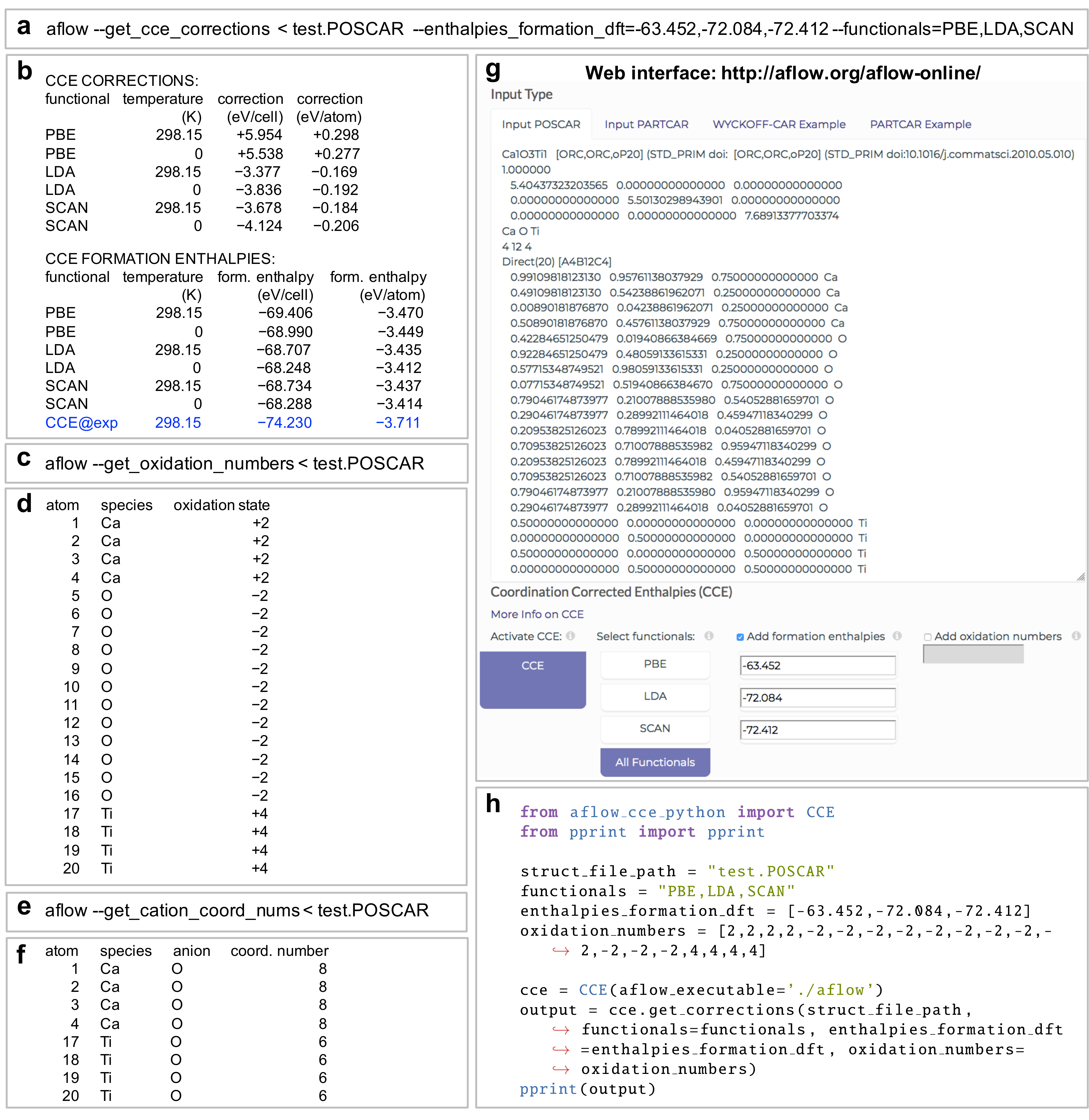}
	\caption{\small \textbf{User interfaces.}
		(\textbf{a}) Example command  --- here for perovskite CaTiO$_3$ --- for the \AFLOW-\CCE\ command line tool.
		The input structure file (here test.{\POSCAR}), precalculated \DFT\ formation enthalpies per cell, and functionals are given via the options \texttt{-}\texttt{-get\_cce\_corrections < test.{\POSCAR}}, \texttt{-}\texttt{-enthalpies\_formation\_dft=-63.452,-72.084,-72.412}, and \texttt{-}\texttt{-functionals=PBE,LDA,SCAN}, respectively.
		The structure can be in any format recognizable by \AFLOW, such as {\VASP\ \POSCAR} \cite{vasp}, Quantum Espresso \cite{quantum_espresso_2009_etal}, {\small FHI}-{\small AIMS} \cite{blum:fhi-aims}, {\small ABINIT} \cite{gonze:abinit_etal}, {\small ELK} \cite{elk} and {\small CIF} \cite{Hall_CIF_1991}.
		Oxidation numbers for all atoms can be provided as a comma separated list as input using \texttt{-}\texttt{-oxidation\_numbers=ox\_num\_1,ox\_num\_2,\dots}.
		(\textbf{b}) When executed, the output includes the \CCE\ corrections and formation enthalpies at both 298.15 and 0~K.
		If no \DFT\ formation enthalpies are given, an estimate for the formation enthalpy at 298.15~K based on experimental values per bond (CCE@exp, blue)~\cite{Friedrich_CCE_2019} is calculated according to Eq.~(\ref{CCE_exp}).
		(\textbf{c/d}) Example command/output when determining oxidation numbers for the structure in test.{\POSCAR}.
		(\textbf{e/f}) Example command/output when determining cation coordination numbers for the structure in test.{\POSCAR}.
		(\textbf{g}) The web interface yields the cation coordination numbers, oxidation numbers, and \CCE\ corrections for the structure provided in the field ``Input {\POSCAR}''.
		If \DFT\ formation enthalpies per cell are provided, the output also includes the \CCE\ formation enthalpies.
		(\textbf{h}) Example Python script using the \AFLOW-\CCE\ Python environment.
		Similar to the command line, \texttt{functionals}, \texttt{enthalpies\_formation\_dft}, and input \texttt{oxidation\_numbers} are optional arguments for the \texttt{get\_corrections} method.
		The results are returned as a dictionary.
        }
	\label{fig1}
\end{figure*}

\noindent
The automated \CCE\ implementation inside \AFLOW\ enables the correction of an extensive library of {ionic} materials that are made available via the \AFLOW\ APIs \cite{curtarolo:art92,curtarolo:art128} and web interfaces \cite{curtarolo:art75}.
The implementation features three ways of user interaction depicted in Fig.~\ref{fig1}: (i) a command line tool, (ii) a web application, and (iii) a Python environment.
The command line tool (Fig.~\ref{fig1}({a}-{f})) provides the \CCE\ corrections and formation enthalpies, (automatically determined) oxidation numbers, and cation coordination numbers for the given structure file that can be in any format recognizable by \AFLOW, such as {\VASP\ \POSCAR} \cite{vasp}, Quantum Espresso \cite{quantum_espresso_2009_etal}, {\small FHI}-{\small AIMS} \cite{blum:fhi-aims}, {\small ABINIT} \cite{gonze:abinit_etal}, {\small ELK} \cite{elk} and {\small CIF} \cite{Hall_CIF_1991}.
Available options
are described in Section ``\CCE\ command line interface''.
The web interface (Fig.~\ref{fig1}({g})) prints the cation coordination numbers,
oxidation numbers and \CCE\ corrections for the selected functionals using the given structure.
The output also includes the \CCE\ formation enthalpies when precalculated \DFT\ values are entered in the designated fields.
The Python environment is distributed with the \AFLOW\ source and can be generated with the command \texttt{aflow --cce
--print=python}.
It connects to the command line functionality and imports the results into a \texttt{CCE} class similar to the Python modules of \AFLOW-{\small SYM} \cite{curtarolo:art135} and \AFLOW-{\small CHULL} \cite{curtarolo:art144}.
An example script leveraging the functionality is depicted in Fig.~\ref{fig1}({h}).
The \CCE\ object has three built-in methods:\\
\texttt{get\_corrections(struct\_file\_path, functionals, enthalpies\_formation\_dft, oxidation\_numbers)},\\
\texttt{get\_oxidation\_numbers(struct\_file\_path)}, and\\
\texttt{get\_cation\_coordination\_numbers(struct\_file\_path})\\
corresponding to the
command line options mentioned in Section ``\CCE\ command line interface''.
Each method requires a path to the input structure file (\texttt{struct\_file\_path}).
For \texttt{get\_corrections}, providing \texttt{functionals}, \DFT\ formation enthalpies (\texttt{enthalpies\_formation\_dft}), and input \texttt{oxidation\_numbers} for all atoms in the structure, are optional arguments.
The results are returned in the form of a Python dictionary.

\vspace{0.7cm}

\noindent
\textbf{\CCE\ command line interface}

\vspace{0.3cm}
\noindent
\texttt{aflow --cce}
\begin{itemize}
\item[-] Prints instructions and example input structure.
\end{itemize}
\noindent
\texttt{aflow --cce={\small STRUCTURE\_FILE\_PATH}}
\begin{itemize}
\item[-] Prints the results of the full \CCE\ analysis, \emph{i.e.} cation coordination numbers, oxidation numbers, and \CCE\ corrections and formation enthalpies, for the given structure. {\small STRUCTURE\_FILE\_PATH} is the path to the structure file. The file can be in any format supported by \AFLOW, \emph{e.g.} {\VASP\ \POSCAR}, {\small QUANTUM ESPRESSO}, {\small AIMS}, {\small ABINIT}, {\small ELK} and {\small CIF}. For {\VASP}, a {\VASP}5 \POSCAR\ is required or, if a {\VASP}4 \POSCAR\ is used, the species must be written on the right side next to the coordinates for each atom just as for the example input structure obtained from \texttt{--cce}.
\end{itemize}
\noindent
\texttt{aflow --get\_cce\_corrections < {\small STRUCTURE\_FILE\_PATH}}
\begin{itemize}
\item[-] Determines the \CCE\ corrections and formation enthalpies for the structure in file {\small STRUCTURE\_FILE\_PATH}.
\end{itemize}
\noindent
\texttt{aflow --get\_oxidation\_number < {\small STRUCTURE\_FILE\_PATH}}
\begin{itemize}
\item[-] Determines the oxidation numbers for the structure in file {\small STRUCTURE\_FILE\_PATH}.
\end{itemize}
\noindent
\texttt{aflow --get\_cation\_coord\_num < {\small STRUCTURE\_FILE\_PATH}}
\begin{itemize}
\item[-] Determines the number of anion neighbors for each cation for the structure in file {\small STRUCTURE\_FILE\_PATH}.
\end{itemize}

\noindent
\textbf{Options for \texttt{--cce={\small STRUCTURE\_FILE\_PATH} and --get\_cce\_corrections < {\small STRUCTURE\_FILE\_PATH}}:}

\vspace{0.3cm}
\noindent
\texttt{--enthalpies\_formation\_dft=enth\_1,enth\_2,...}
\begin{itemize}
\item[-] \texttt{enth\_1,enth\_2,...} is a comma separated list for precalculated \DFT\ formation enthalpies. They are assumed to be: (i) negative for compounds lower in enthalpy than the elements, (ii) in eV/cell. Currently, corrections are available for \PBE, \LDA, and \SCAN.
\end{itemize}
\noindent
\texttt{--functionals=func\_1,func\_2,func\_3}
\begin{itemize}
\item[-] \texttt{func\_1,func\_2,func\_3} is a comma separated list of functionals for which corrections should be returned. If used together with \texttt{--enthalpies\_formation\_dft}, the functionals must be in the same sequence as the \DFT\ formation enthalpies they correspond to. Available functionals are: (i) \PBE, (ii) \LDA, and (iii) \SCAN. Default: \PBE\ (if only one \DFT\ formation enthalpy is provided).
\end{itemize}
\noindent
\texttt{--oxidation\_numbers=ox\_num\_1,ox\_num\_2,...}
\begin{itemize}
\item[-] \texttt{ox\_num\_1,ox\_num\_2,...} is a comma separated list of oxidation numbers. It is assumed that: (i) one is provided for each atom of the structure and (ii) they are in the same sequence as the corresponding atoms in the provided structure file.
\end{itemize}

\noindent\textbf{General option}

\vspace{0.3cm}
\noindent
\texttt{--print=out|json}
\begin{itemize}
\item[-] Obtains output in table format (\texttt{--print=out}) or as {\small JSON} (\texttt{--print=json}). Default: \texttt{out}.
\end{itemize}

\noindent
\textbf{Structural analysis.}
For evaluating the number of cation-anion bonds (cation coordination numbers), first the (main) anion species of the system is determined as the one with the highest Allen \underline{e}lectro\underline{n}egativity (EN) \cite{Allen_electronegativity_1989,Mann_JACS_2000,Mann_JACS_2000_2}.
A check is performed whether the material is a multi-anion system \cite{Kageyama_NCOM_2018}, \emph{i.e.} whether atoms of a type other than the main anion species are only bound to atoms of lower EN or its own type.
If such atoms are found, they are designated as additional anions.
This is for instance the case for N in HfTaNO$_3$, where O is the main anion.
Note that in some compounds certain species can occur both as anion and as cation: in ammonium-nitrate (NH$_4$NO$_3$) for instance, N occurs both in $-3$ and $+5$ oxidation states depending on its neighbors.\par

Subsequently, the number of anion neighbors for each cation is determined.
For this bonding analysis, the nearest neighbor distance is obtained for each species.
A bonding cutoff is set by adding a tolerance of 0.5~\AA\ in accordance with Ref.~\onlinecite{Friedrich_CCE_2019}.
Then, all anion neighbors between the species selective minimum distance and the cutoff are counted.
For the multi-anion analysis, the tolerance is reduced to 0.4~\AA\ since for larger values, different anions of systems known to be multi-anion compounds would be detected as being bonded.
For instance, in HfTaNO$_3$, if the tolerance is not reduced, N and O would be detected as being neighbors and hence nitrogen would not be identified as an anion.
\par

When oxygen is found as an anion, the O-O distances in the system are determined to detect per- and superoxides. The following scenarios can occur: (i) the O-O bond is longer than 1.6~\AA\ indicating an oxide (O$^{2-}$ ion), (ii) the bond length is between 1.4 and 1.6~\AA\ (peroxide), (iii) the bond length lies between 1.3 and 1.4~\AA, (superoxide), and (iv) the bond length is shorter than 1.3~\AA, \emph{i.e.} the structure may contain molecular oxygen the enthalpy of which is not correctable within \CCE.
For certain special cases such as alkali metal sesquioxides, several of the above scenarios can be fulfilled simultaneously and the implementation will then treat the system as incorporating multiple different oxygen ions.
The separation of the different oxide types by bond length is based on the study of the relaxed structures of Li$_2$O$_2$, Na$_2$O$_2$, K$_2$O$_2$, SrO$_2$, BaO$_2$ (peroxides), NaO$_2$, KO$_2$, CsO$_2$ (superoxides), and O$_2$ \cite{Friedrich_CCE_2019}.
The number of (su-)peroxide bonds is determined as half the number of (su-)peroxide O atoms.

\setlength\tabcolsep{11pt}
\begin{table*}[htb!]
	\begin{center}
		\caption{\small \textbf{Electronegativities and oxidation numbers.}
                  \label{tab_1}
        Allen ENs \cite{Allen_electronegativity_1989,Mann_JACS_2000,Mann_JACS_2000_2}, as well as preferred and all known oxidation numbers according to Ref.~\onlinecite{PSE_Wiley_2012} with additions, used in the \CCE\ implementation.
        For Cr, the most preferred value is listed first.
		}
			\begin{tabular}{@{}lccc|lccc@{}}
				\hline
				element & Allen & \multicolumn{2}{c}{oxidation numbers} & element & Allen & \multicolumn{2}{c}{oxidation numbers} \\
				 & EN & preferred & all &  & EN & preferred & all \\
				\hline
H	&	2.3	&	+1	&	+1,$-$1	&	Rh	&	1.56	&	+3,+1	&	+5,+4,+3,+2,+1,0	\\
He	&	4.16	&	-	&	-	&	Pd	&	1.58	&	+2	&	+4,+2,0	\\
Li	&	0.912	&	+1	&	+1	&	Ag	&	1.87	&	+1	&	+2,+1	\\
Be	&	1.576	&	+2	&	+2	&	Cd	&	1.52	&	+2	&	+2	\\
B	&	2.051	&	+3	&	+3	&	In	&	1.656	&	+3	&	+3	\\
C	&	2.544	&	+4,$-$4	&	+4,+2,$-$4	&	Sn	&	1.824	&	+4,+2	&	+4,+2	\\
N	&	3.066	&	$-$3	&	+5,+4,+3,+2,$-$3	&	Sb	&	1.984	&	+3	&	+5,+3,$-$3	\\
O	&	3.61	&	$-$2	&	$-$0.5,$-$1,$-$2	&	Te	&	2.158	&	+4	&	+6,+4,$-$2	\\
F	&	4.193	&	$-$1	&	$-$1	&	I	&	2.359	&	$-$1	&	+7,+5,+1,$-$1	\\
Ne	&	4.787	&	-	&	-	&	Xe	&	2.582	&	-	&	+8,+6,+4,+2	\\
Na	&	0.869	&	+1	&	+1	&	Cs	&	0.659	&	+1	&	+1	\\
Mg	&	1.293	&	+2	&	+2	&	Ba	&	0.881	&	+2	&	+2	\\
Al	&	1.613	&	+3	&	+3	&	La\footnotemark	&	1.09	&	+3	&	+3	\\
Si	&	1.916	&	+4	&	+4,$-$4	&	Ce\footnotemark[1]	&	1.09	&	+3	&	+4,+3	\\
P	&	2.253	&	+5	&	+5,+3,$-$3	&	Pr\footnotemark[1]	&	1.09	&	+3	&	+4,+3	\\
S	&	2.589	&	+6	&	+6,+4,+2,$-$2	&	Nd\footnotemark[1]	&	1.09	&	+3	&	+3	\\
Cl	&	2.869	&	$-$1	&	+7,+5,+3,+1,$-$1	&	Pm\footnotemark[1]	&	1.09	&	+3	&	+3	\\
Ar	&	3.242	&	-	&	-	&	Sm\footnotemark[1]	&	1.09	&	+3	&	+3,+2	\\
K	&	0.734	&	+1	&	+1	&	Eu\footnotemark[1]	&	1.09	&	+3	&	+3,+2	\\
Ca	&	1.034	&	+2	&	+2	&	Gd\footnotemark[1]	&	1.09	&	+3	&	+3	\\
Sc	&	1.19	&	+3	&	+3	&	Tb\footnotemark[1]	&	1.09	&	+3	&	+4,+3	\\
Ti	&	1.38	&	+4	&	+4,+3,+2	&	Dy\footnotemark[1]	&	1.09	&	+3	&	+3	\\
V	&	1.53	&	+5	&	+5,+4,+3,+2,0	&	Ho\footnotemark[1]	&	1.09	&	+3	&	+3	\\
Cr	&	1.65	&	+3,+6	&	+6,+3,+2,0	&	Er\footnotemark[1]	&	1.09	&	+3	&	+3	\\
Mn	&	1.75	&	+2	&	+7,+6,+4,+3,+2,0,$-$1	&	Tm\footnotemark[1]	&	1.09	&	+3	&	+3,+2	\\
Fe	&	1.8	&	+3,+2	&	+6,+3,+2,0,$-$2	&	Yb\footnotemark[1]	&	1.09	&	+3	&	+3,+2	\\
Co	&	1.84	&	+2	&	+3,+2,0,$-$1	&	Lu	&	1.09	&	+3	&	+3	\\
Ni	&	1.88	&	+2	&	+3,+2,0	&	Hf	&	1.16	&	+4	&	+4,+3	\\
Cu	&	1.85	&	+2,+1	&	+2,+1	&	Ta	&	1.34	&	+5	&	+5,+3	\\
Zn	&	1.59	&	+2	&	+2	&	W	&	1.47	&	+6	&	+6,+5,+4,+3,+2,0	\\
Ga	&	1.756	&	+3	&	+3	&	Re	&	1.6	&	+7	&	+7,+6,+4,+2,$-$1	\\
Ge	&	1.994	&	+4	&	+4	&	Os	&	1.65	&	+4	&	+8,+6,+4,+3,+2,0,$-$2	\\
As	&	2.211	&	+3	&	+5,+3,$-$3	&	Ir	&	1.68	&	+4,+1	&	+6,+4,+3,+2,+1,0,$-$1	\\
Se	&	2.424	&	+4	&	+6,+4,$-$2	&	Pt	&	1.72	&	+4,+2	&	+4,+2,0	\\
Br	&	2.685	&	$-$1	&	+7,+5,+3,+1,$-$1	&	Au	&	1.92	&	+3	&	+3,+1	\\
Kr	&	2.966	&	+2	&	+2	&	Hg	&	1.76	&	+2	&	+2,+1	\\
Rb	&	0.706	&	+1	&	+1	&	Tl	&	1.789	&	+1	&	+3,+1	\\
Sr	&	0.963	&	+2	&	+2	&	Pb	&	1.854	&	+2	&	+4,+2	\\
Y	&	1.12	&	+3	&	+3	&	Bi	&	2.01	&	+3	&	+5,+3	\\
Zr	&	1.32	&	+4	&	+4,+3	&	Po	&	2.19	&	+4	&	+6,+4,+2	\\
Nb	&	1.41	&	+5	&	+5,+3,+2	&	At	&	2.39	&	$-$1	&	+7,+5,+3,+1,$-$1	\\
Mo	&	1.47	&	+6	&	+6,+5,+4,+3,+2,0	&	Rn	&	2.6	&	+2	&	+2	\\
Tc	&	1.51	&	+7	&	+7	&	Fr	&	0.67	&	+1	&	+1	\\
Ru	&	1.54	&	+4,+3	&	+8,+6,+4,+3,+2,0,$-$2	&	Ra	&	0.89	&	+2	&	+2	\\
				\hline
			\end{tabular}
			\footnotetext[1]{Since there are no available Allen electronegativities for La-Yb, the value for Lu is used as these elements are usually very similar. This is confirmed by the Allred and Rochow electronegativities that are very similar for all lanthanides \cite{Allred_Elneg_1958}.}
	\end{center}
\end{table*}

\ \\

\noindent
\textbf{Determination of oxidation numbers.}
The default method to automatically determine oxidation numbers is based on Allen ENs \cite{Allen_electronegativity_1989,Mann_JACS_2000,Mann_JACS_2000_2}.
This choice is in accordance with \underline{I}nternational \underline{U}nion of \underline{P}ure and \underline{A}pplied \underline{C}hemistry (IUPAC) recommendations \cite{Karen_IUPAC_2014,Karen_IUPAC_2016} and also conforms with our own tests that this EN scale yields the most reliable oxidation numbers when compared to other scales such as Refs.~\onlinecite{Pauling_JACS_1932,Allred_Elneg_1958}.
Table~\ref{tab_1} lists the EN values together with the preferred and all known oxidation numbers for the elements according to Ref.~\onlinecite{PSE_Wiley_2012}, along with a few additions deemed necessary during the test of the implementation.
The separation into preferred and all known oxidation numbers is motivated by the finding that in compounds with more than two species, elements tend to occur only in the preferred oxidation states \cite{Friedrich_CCE_2019}.
Missing oxidation states will be added in future releases as needed.\par

\begin{figure}[ht!]
	\centering
	\includegraphics[width=0.99\columnwidth]{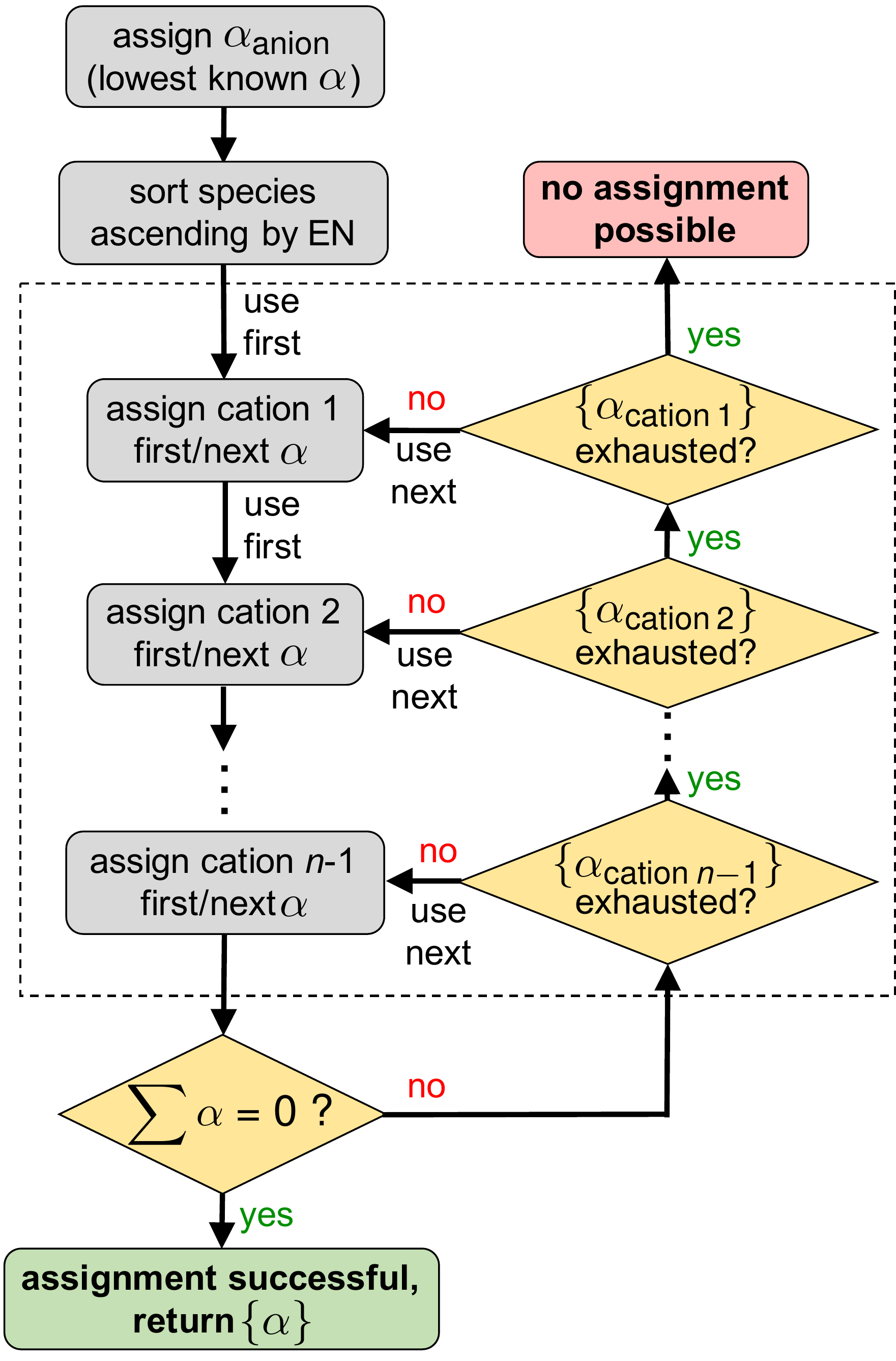}
	\caption{\small \textbf{Oxidation number algorithm.}
		Schematic representation of the algorithm to determine oxidation numbers $\alpha$ of a compound with $n$-species.
		The part of the algorithm for determining cation oxidation numbers (inside the black dashed box) is first applied making use of the preferred oxidation numbers for all species.
		If no successful assignment is achieved, it is employed a second time after checking for mixed-valence compounds using all known oxidation states (Table~\ref{tab_1}).
		The oxidation numbers of more electronegative cation species are iterated faster than the more electropositive ones.
		Three dots indicate proceeding equivalently for all further cation species.
		For multi-anion systems, atoms identified as additional anions during the structural analysis are excluded when assigning cation oxidation numbers.
        }
	\label{fig2}
\end{figure}

The algorithm (Fig.~\ref{fig2}) starts by assigning the anion oxidation numbers.
For all anion atoms (main anion species and anions from multi-anion analysis) the lowest (most negative) oxidation number known for this species is assigned.
If the atom was found to belong to a peroxide (superoxide) ion in the structural analysis, the oxidation number is changed to $-1$ ($-0.5$).\par

The set of possible cation oxidation states is first restricted to the preferred values for each species.
All cations are then assigned the first (usually most positive) preferred oxidation number for their species.
The only exception is Cr for which $+3$ is the first choice (Table~\ref{tab_1}).
After this initial assignment, the sum over all oxidation numbers is evaluated and --- if it is zero ---
the assignment is considered successful.
Otherwise, the algorithm proceeds by changing the preferred oxidation numbers according to EN:
while checking the sum for each choice, the oxidation states of more electronegative (higher EN) cation species are changed to the next preferred value before the more electropositive (lower EN) ones.
It is expected that more electropositive elements occur in higher oxidation states.

If all EN-directed choices of preferred oxidation numbers are exhausted without successful assignment, the system is checked for mixed-valence.
For these special cases, the oxidation numbers are set explicitly.
Currently, this scenario includes Sb$_2$O$_4$, Pb$_3$O$_4$, Fe$_3$O$_4$, Mn$_3$O$_4$, Co$_3$O$_4$, Ti-O Magn\'{e}li phases, and alkali-metal sesquioxides.
If still no successful assignment is achieved, the part of the algorithm for determining cation oxidation numbers (inside the black dashed box in Fig.~\ref{fig2}) is repeated with all known oxidation numbers for all cation species.
The scheme has been successfully tested on a large number of compounds, including oxides, fluorides, chlorides, and nitrides.
The algorithm might not be particularly suited for organic compounds for which the oxidation state of C depends on the functional group.
Such materials are presently beyond the scope of \AFLOW-\CCE.

For oxides, the oxidation numbers can also be determined from Bader
charges \cite{Henkelman_CMS_2006}, which are compared to the averaged template values of the binary fit set for the respective functional.
The formal oxidation number is assigned according to the closest template value.
However, this scheme shows systematic difficulties in assigning the correct oxidation numbers for several species in certain oxidation states such as Ti$^{4+}$, V$^{5+}$, Fe$^{2+}$ and Fe$^{3+}$, for which error handling procedures have been implemented.
This method is only invoked when specifically requested via the
setting \texttt{DEFAULT\_CCE\_OX\_METHOD=2} in the \texttt{.aflow.rc}
setup file of \AFLOW.

Finally, the user can also provide the oxidation numbers for all atoms
as a comma separated list as input (option
\texttt{--oxidation\_numbers=} in the Section ``\CCE\ command line interface'').

\begin{figure}[ht!]
	\centering
	\includegraphics[width=0.99\columnwidth]{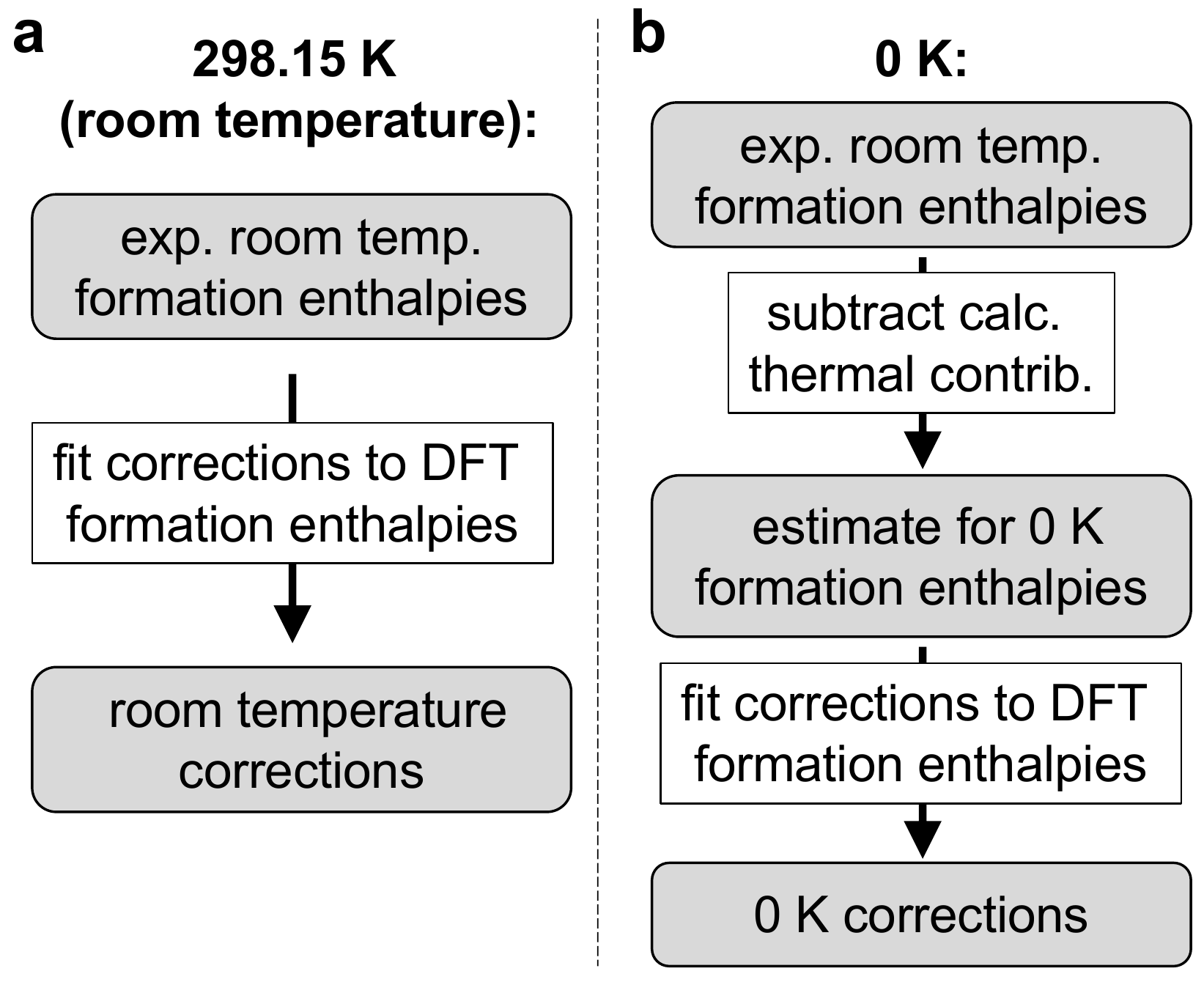}
	\caption{\small \textbf{Corrections for different temperatures.}
		(\textbf{a}) For 298.15~K, the \CCE\ corrections to the DFT formation enthalpies are fitted to experimental room temperature formation enthalpies resulting in room temperature corrections.
		(\textbf{b}) For 0~K, first the thermal contribution deduced from a quasi-harmonic Debye model \cite{curtarolo:art96} is subtracted from experimental values, resulting in estimates for 0~K formation enthalpies.
		\CCE\ corrections fitted to these values yield 0~K corrections.
        }
	\label{fig3}
\end{figure}

\noindent
\textbf{Inclusion of temperature effects.} \label{temp_eff}
After the determination of the oxidation numbers, the cation-anion and cation oxidation state specific \CCE\ corrections per bond $\delta H^{T,A^{+\alpha}}_{A-Y}$ (Eq.~(\ref{fit_CCE})) can be assigned.
As outlined in Ref.~\onlinecite{Friedrich_CCE_2019}, vibrational (zero-point and thermal) contributions to the formation enthalpy do not need to be calculated explicitly since they can be parameterized per bond and thus implicitly included into the corrections.
Compared to when the vibrational contribution was explicitly included for the fit and test sets, the \MAE\ of the corrected results increased by at most 1~meV/atom.
This is negligible considering that the \CCE\ \MAE\ is on the order of 30~meV/atom.
Temperature effects are thus included in the corrections according to Fig.~\ref{fig3}({a}):
The \CCE\ corrections to \DFT\ formation enthalpies are fitted to experimental room temperature formation enthalpies resulting in room temperature corrections.
When these are applied according to Eq.~(\ref{apply_CCE}), a direct estimate of the room temperature formation enthalpy is obtained.\par

For 0~K (Fig.~\ref{fig3}({b})), one first subtracts the calculated thermal contribution, deduced from a quasi-harmonic Debye model \cite{curtarolo:art96} according to Ref.~\cite{Friedrich_CCE_2019}, from the experimental formation enthalpy for each functional.
This gives a good estimate for the (experimental) 0~K formation enthalpy.
Then, the \CCE\ corrections to the DFT formation enthalpies are fitted to these values yielding 0~K corrections.
The approach does not capture any phase change of the elemental references from 0~K to room temperature.
However, this is a rare event that occurs on an energy scale \textit{below} room temperature, which is smaller than the \CCE\ error.\par

\begin{figure}[ht!]
	\centering
	\includegraphics[width=0.99\columnwidth]{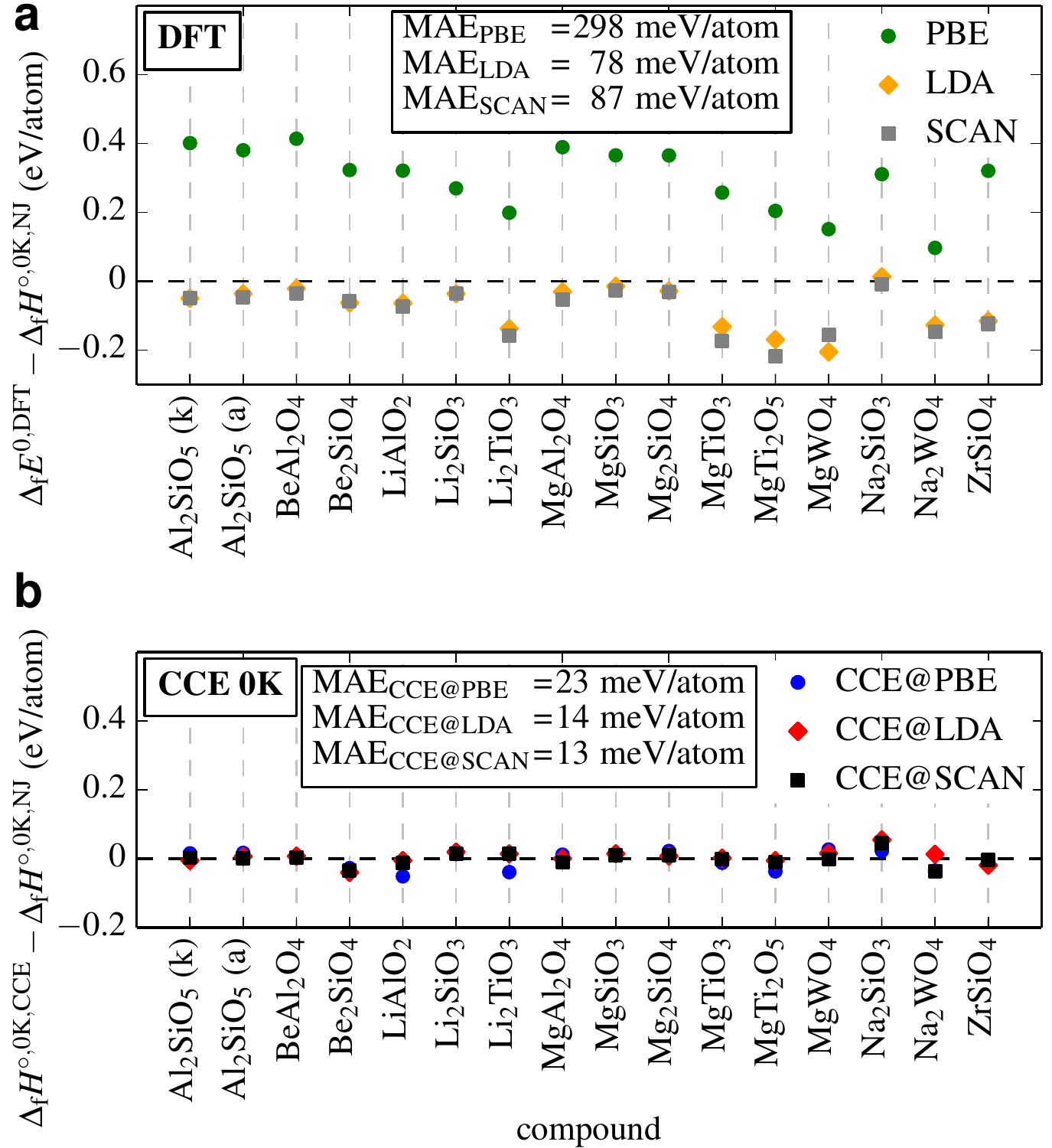}
	\caption{\small \textbf{Validating 0~K predictions.}
		Deviations of (\textbf{a}) \DFT\ formation enthalpies and (\textbf{b}) \CCE\ 0~K predictions from experimental 0~K formation enthalpies from {\small NIST-JANAF} \cite{Chase_NIST_JANAF_thermochem_tables_1998}.
		For Al$_2$SiO$_5$, the results for both the kyanite (k) and andalusite (a) structures are depicted.
        }
	\label{fig4}
\end{figure}

To test the accuracy of the predicted 0~K formation enthalpies, they are compared to available tabulated values.
Figure~\ref{fig4}({a}) shows the deviations between plain \DFT\ results and 0~K formation enthalpies from the {\small \underline{N}IST-\underline{J}ANAF} (\NJ) thermochemical tables \cite{Chase_NIST_JANAF_thermochem_tables_1998} for 16 ternary oxides.
Typical large errors are indicated by the {\MAE}s of 298, 78 and 87~meV/atom for \PBE, \LDA\ and \SCAN, respectively.
When corrected by \CCE, the \DFT\ results are drastically improved (Fig.~\ref{fig4}({b})) with mean errors reduced to 23, 14 and 13~meV/atom, validating our 0~K approach.

As a future development, the temperature dependence can be implemented as a continuous variable, since it can be parameterized per bond and the thermal contributions at any temperature can be computed from the quasiharmonic Debye model \cite{curtarolo:art96} for the fit set.
{This ansatz, and other approaches directly targeting the Gibbs free energy \cite{Bartel_NCOM_2018} to include finite temperature effects, pave the way to move beyond stability predictions based only on enthalpies, which are crucial for \emph{e.g.} high-entropy materials \cite{curtarolo:art99,curtarolo:art140}.}\par

The corrections are finally used to calculate the total \CCE\ corrections and the \CCE\ formation enthalpies according to Eq.~(\ref{apply_CCE}) for 298.15 and 0~K for all functionals selected.
If needed, corrections for (su-)peroxides are added according to Ref.~\cite{Friedrich_CCE_2019} with the number of respective O-O bonds obtained from the structural analysis.
If no precalculated \DFT\ formation enthalpies are provided, an estimate for the formation enthalpy at 298.15~K based on experimental values per bond (CCE@exp)~\cite{Friedrich_CCE_2019} is calculated according to Eq.~(\ref{CCE_exp}).

\onecolumngrid

\LTcapwidth=\textwidth

\newcommand\hdl[1]{\multicolumn{1}{l}{#1}}
\newcommand\hdr[1]{\multicolumn{1}{r}{#1}}
\newcommand\hd[1]{\multicolumn{1}{c}{#1}}
\newcommand\zz[1]{^{#1}\!\!}
\setlength\tabcolsep{11pt}
\begin{longtable*}{@{}lr*{1}{D..{\tablecolumnspacing}}*{1}{D..{\tablecolumnspacing}}*{1}{D..{\tablecolumnspacing}}|lr*{1}{D..{\tablecolumnspacing}}*{1}{D..{\tablecolumnspacing}}*{1}{D..{\tablecolumnspacing}}@{}}
  \caption{\small \textbf{\CCE\ corrections at 0~K for oxides.}
    Corrections per bond $\delta H^{\mathrm{0K},A^{+\alpha}}_{A-Y}$ of the \CCE\ method for each cation species $A$ in oxidation states $+\alpha$ for 0~K obtained from binary oxides.
    The corrections for Si and Ti in oxidation state +4 are obtained from $\alpha$-quartz (\AFLOW\ label \texttt{A2B\_hP9\_152\_c\_a}~\citePROTOS) and rutile (\AFLOW\ label \texttt{A2B\_tP6\_136\_f\_a}~\citePROTOS), respectively.
    The corrections in the last line are for (su-)peroxides according to the approach outlined in Ref.~\onlinecite{Friedrich_CCE_2019}.
    All corrections are in eV/bond.
  }\label{tab_2}\\
  \hline
  cation & \hd{$+\alpha$} & \multicolumn{3}{c}{$\delta H^{\mathrm{0K},A^{+\alpha}}_{A-Y}$} & cation & \hd{$+\alpha$} & \multicolumn{3}{c}{$\delta H^{\mathrm{0K},A^{+\alpha}}_{A-Y}$} \\
  species $A$ & & \hd{\PBE} & \hd{\LDA} & \hd{\hspace{0.1cm} \SCAN} & species $A$ & & \hd{\PBE} & \hd{\LDA} & \hd{\hspace{0.5cm} \SCAN} \\
  \hline
  \endfirsthead

  \multicolumn{10}{c}
  {{\tablename\ \thetable{}. (\textit{continued})}} \\
  \hline
  cation & \hd{$+\alpha$} & \multicolumn{3}{c}{$\delta H^{\mathrm{0K},A^{+\alpha}}_{A-Y}$} & cation & \hd{$+\alpha$} & \multicolumn{3}{c}{$\delta H^{\mathrm{0K},A^{+\alpha}}_{A-Y}$} \\
  species $A$ & & \hd{\PBE} & \hd{\LDA} & \hd{\hspace{0.1cm} \SCAN} & species $A$ & & \hd{\PBE} & \hd{\LDA} & \hd{\hspace{0.5cm} \SCAN} \\
  \hline
	\endhead
	Li	&	+1	&	 0.0704	&	-0.0223	&	-0.0186 &	Sr	&	+2	&	 0.1048	&	-0.0232	&	-0.0193	\\
	Be	&	+2	&	 0.1875	&	 0.0000 &	-0.0023	&	Y 	&	+3	&	 0.1304	&	-0.0280	&	-0.0543	\\
	B 	&	+3	&	 0.1825	&	-0.0835	&	-0.0612	&	Zr	&	+4	&	 0.1320	&	-0.0530	&	-0.0709	\\
	Na	&	+1	&	 0.0776	&	-0.0096	&	-0.0168 &	Nb	&	+2	&	 0.0533	&	-0.1328	&	-0.0910	\\
	Mg	&	+2	&	 0.1272	&	-0.0042	&	-0.0090 &	Mo	&	+4	&	 0.0215	&	-0.1927 &	-0.1137 \\
	Al	&	+3	&	 0.1778	&	-0.0168 &	-0.0222 &	Mo	&	+6	&	-0.0603	&	-0.3575	&	-0.2718 \\
	Si ($\alpha$-qua.)	&	+4	&	 0.2380	&	-0.0390	&	-0.0368	&	Ru	&	+4	&	-0.0115	&	-0.2133	&	-0.1192	\\
	K 	&	+1	&	 0.0803	&	-0.0301	&	-0.0071 &	Rh	&	+3	&	 0.0065	&	-0.1415	&	-0.0631	\\
	Ca	&	+2	&	 0.1002	&	-0.0395	&	-0.0280	&	Pd	&	+2	&	 0.0548	&	-0.0830	&	-0.0280	\\
	Sc	&	+3	&	 0.1541	&	-0.0166	&	-0.0338	&	Ag	&	+1	&	-0.0070	&	-0.0538 &	-0.0645	\\
	Ti	&	+2	&	 0.1067	&	-0.0738	&	-0.0331	&	Cd	&	+2	&	 0.1013	&	 0.0130	&	 0.0023	\\
	Ti	&	+3	&	 0.0905	&	-0.0936	&	-0.0767	&	In	&	+3	&	 0.1303	&	-0.0222	&	-0.0186 \\
	Ti (rut.)	&	+4	&	 0.0972	&	-0.1072	&	-0.1345	&	Sn	&	+2	&	 0.0650	&	-0.0665	&	-0.0158	\\
	V 	&	+2	&	 0.2620	&	 0.1152	&	 0.1547 &	Sn	&	+4	&	 0.1433	&	-0.0540	&	-0.0237	\\
	V 	&	+3	&	 0.0918	&	-0.0734	&	-0.0600	&	Sb	&	+3	&	 0.1153	&	-0.1212 &	-0.0267 \\
	V 	&	+4	&	 0.0375	&	-0.1598	&	-0.1637 &	Sb	&	+5	&	 0.0970	&	-0.1418	&	-0.0669	\\
	V 	&	+5	&	-0.0189	&	-0.2248	&	-0.2307 &	Te	&	+4	&	 0.0558	&	-0.2123 &	-0.0970 \\
	Cr	&	+6	&	-0.1443	&	-0.3210	&	-0.2968	&	Cs	&	+1	&	 0.1008	&	-0.0583	&	-0.0060	\\
	Cr	&	+3	&	 0.1473	&	 0.0391	&	-0.0247	&	Ba	&	+2	&	 0.1165	&	 0.0075 &	 0.0020 \\
	Mn	&	+2	&	 0.2513	&	 0.2693	&	-0.0340 &	Hf	&	+4	&	 0.1566	&	-0.0353 &	-0.0326	\\
	Mn	&	+4	&	 0.0523	&	-0.1030 &	-0.1667 &	W 	&	+4	&	 0.0512	&	-0.1648	&	-0.0543	\\
	Fe	&	+2	&	 0.1728	&	 0.1287	&	 0.0143	&	W 	&	+6	&	-0.0025	&	-0.2165	&	-0.1448	\\
	Fe	&	+3	&	 0.1586	&	 0.0055	&	-0.0718	&	Re	&	+4	&	 0.0845	&	-0.1302	&	 0.0155	\\
	Co	&	+2	&	 0.2373	&	 0.1645	&	 0.1193	&	Re	&	+6	&	-0.0803	&	-0.3125	&	-0.1682	\\
	Ni	&	+2	&	 0.2537	&	 0.1512	&	 0.1955 &	Os	&	+4	&	 0.0570	&	-0.1498	&	-0.0040	\\
	Cu	&	+1	&	 0.1295	&	 0.0318 &	 0.0618 &	Os	&	+8	&	-0.2295	&	-0.3920	&	-0.2880	\\
	Cu	&	+2	&	 0.0973	&	-0.0308	&	 0.0020	&	Ir	&	+4	&	 0.0157	&	-0.1868	&	 0.0135	\\
	Zn	&	+2	&	 0.1803	&	 0.0368	&	 0.0398	&	Hg	&	+2	&	 0.1525	&	-0.0865	&	 0.0380	\\
	Ga	&	+3	&	 0.1925	&	-0.0022 &	 0.0289	&	Tl	&	+1	&	-0.0053	&	-0.0660	&	-0.0605 \\
	Ge	&	+4	&	 0.1895	&	-0.0462	&	 0.0290	&	Tl	&	+3	&	 0.0518	&	-0.0962	&	-0.0166	\\
	As	&	+5	&	 0.1919	&	-0.0752	&	 0.0092	&	Pb	&	+2	&	 0.0033	&	-0.1093	&	-0.0550	\\
	Se	&	+4	&	 0.0657	&	-0.2397	&	-0.1083	&	Pb	&	+4	&	 0.0545	&	-0.1285	&	-0.0283	\\
	Rb	&	+1	&	 0.0940	&	-0.0235	&	 0.0031	&	Bi	&	+3	&	-0.0276	&	-0.1778	&	-0.0379	\\
	O 	&	$-$1 	&	-0.0856	&	-0.1110	&	 0.2476	&	O 	&	$-\frac{1}{2}$ 	&	-0.5435	&	-0.2697	&	-0.0468	\\
	\hline
\end{longtable*}

\
\setlength\tabcolsep{11pt}
\begin{longtable*}{@{}lr*{1}{D..{\tablecolumnspacing}}|lr*{1}{D..{\tablecolumnspacing}}|lr*{1}{D..{\tablecolumnspacing}}@{}}
  \caption{\small \textbf{\CCE@exp\ corrections at 298.15~K for oxides.}
    Corrections per bond $\delta H^{\mathrm{298.15K},A^{+\alpha}}_{A-Y,\mathrm{exp}}$ of the \CCE@exp method for each cation species $A$ in oxidation states $+\alpha$ for 298.15~K obtained from binary oxides.
    The corrections for Si and Ti in oxidation state +4 are obtained from $\alpha$-quartz (\AFLOW\ label \texttt{A2B\_hP9\_152\_c\_a}~\citePROTOS) and rutile (\AFLOW\ label \texttt{A2B\_tP6\_136\_f\_a}~\citePROTOS), respectively.
    The corrections in the last line are for (su-)peroxides according to the approach outlined in Ref.~\onlinecite{Friedrich_CCE_2019}.
    All corrections are in eV/bond.
  }\label{tab_3}\\
  \hline
  cation & \hd{$+\alpha$} & \multicolumn{1}{c}{$\delta H^{\mathrm{298.15K},A^{+\alpha}}_{A-Y,\mathrm{exp}}$} & cation & \hd{$+\alpha$} & \multicolumn{1}{c}{$\delta H^{\mathrm{298.15K},A^{+\alpha}}_{A-Y,\mathrm{exp}}$} & cation & \hd{$+\alpha$} & \multicolumn{1}{c}{$\delta H^{\mathrm{298.15K},A^{+\alpha}}_{A-Y,\mathrm{exp}}$} \\
  species $A$ & & & species $A$ & & & species $A$ & & \\
  \hline
  \endfirsthead

  \multicolumn{9}{c}
  {{\tablename\ \thetable{}. (\textit{continued})}} \\
  \hline
  cation & \hd{$+\alpha$} & \multicolumn{1}{c}{$\delta H^{\mathrm{298.15K},A^{+\alpha}}_{A-Y,\mathrm{exp}}$} & cation & \hd{$+\alpha$} & \multicolumn{1}{c}{$\delta H^{\mathrm{298.15K},A^{+\alpha}}_{A-Y,\mathrm{exp}}$} & cation & \hd{$+\alpha$} & \multicolumn{1}{c}{$\delta H^{\mathrm{298.15K},A^{+\alpha}}_{A-Y,\mathrm{exp}}$} \\
  species $A$ & & & species $A$ & & & species $A$ & & \\
  \hline
	\endhead
	Li	&	+1	&	-0.7746	&	Fe	&	+3	&	-0.7112	&	In	&	+3	&	-0.7997	\\
	Be	&	+2	&	-1.5790	&	Co	&	+2	&	-0.4107	&	Sn	&	+2	&	-0.7405	\\
	B 	&	+3	&	-2.1998	&	Ni	&	+2	&	-0.4140	&	Sn	&	+4	&	-1.0033	\\
	Na	&	+1	&	-0.5415	&	Cu	&	+1	&	-0.4423	&	Sb	&	+3	&	-1.2370	\\
	Mg	&	+2	&	-1.0392	&	Cu	&	+2	&	-0.4043	&	Sb	&	+5	&	-0.8394	\\
	Al	&	+3	&	-1.4473	&	Zn	&	+2	&	-0.9083	&	Te	&	+4	&	-0.8380	\\
	Si ($\alpha$-qua.)	&	+4	&	-2.3603	&	Ga	&	+3	&	-1.1288	&	Cs	&	+1	&	-0.5977	\\
	K 	&	+1	&	-0.4705	&	Ge	&	+4	&	-1.0018	&	Ba	&	+2	&	-0.9468	\\
	Ca	&	+2	&	-1.0967	&	As	&	+5	&	-0.9536	&	Hf	&	+4	&	-1.6949	\\
	Sc	&	+3	&	-1.6482	&	Se	&	+4	&	-0.7777	&	W 	&	+4	&	-1.0183	\\
	Ti	&	+2	&	-1.1719	&	Rb	&	+1	&	-0.4390	&	W 	&	+6	&	-1.4557	\\
	Ti	&	+3	&	-1.3136	&	Sr	&	+2	&	-1.0227	&	Re	&	+4	&	-0.7755	\\
	Ti (rut.)	&	+4	&	-1.6307	&	Y 	&	+3	&	-1.6453	&	Re	&	+6	&	-1.0177	\\
	V 	&	+2	&	-0.7458	&	Zr	&	+4	&	-1.6249	&	Os	&	+4	&	-0.5088	\\
	V 	&	+3	&	-1.0527	&	Nb	&	+2	&	-1.0875	&	Os	&	+8	&	-1.0200	\\
	V 	&	+4	&	-1.2330	&	Mo	&	+4	&	-1.0155	&	Ir	&	+4	&	-0.4192	\\
	V 	&	+5	&	-1.6067	&	Mo	&	+6	&	-1.9308	&	Hg	&	+2	&	-0.4705	\\
	Cr	&	+3	&	-0.9800	&	Ru	&	+4	&	-0.5268	&	Tl	&	+1	&	-0.2892	\\
	Cr	&	+6	&	-1.5210	&	Rh	&	+3	&	-0.3072	&	Tl	&	+3	&	-0.3372	\\
	Mn	&	+2	&	-0.6648	&	Pd	&	+2	&	-0.2993	&	Pb	&	+2	&	-0.5685	\\
	Mn	&	+4	&	-0.8998	&	Ag	&	+1	&	-0.0805	&	Pb	&	+4	&	-0.4742	\\
	Fe	&	+2	&	-0.4700	&	Cd	&	+2	&	-0.4463	&	Bi	&	+3	&	-0.5915	\\
	O 	&	$-$1 	&	2.7256		&	O 	&	$-\frac{1}{2}$ 	&	1.7560	& & & \\
	\hline
\end{longtable*}

\twocolumngrid

\section{Conclusions} \label{conclusions}
\noindent
We have presented our implementation of the coordination corrected enthalpies (\CCE) method into \AFLOW\ for automated  correction of \DFT\ formation enthalpies.
\AFLOW-\CCE\ provides a universal tool to obtain highly accurate formation enthalpies for {ionic} materials with a typical mean absolute error close to the room temperature thermal energy, \emph{i.e.} $\approx$ 25~meV/atom \cite{Friedrich_CCE_2019}.
It interoperates with the existing functionality of \AFLOW\ and features a command line tool, a web interface, and a Python environment.
Additionally, the \AFLOW-{\small CHULL} module will be updated with the \CCE\ formation enthalpies
where appropriate \cite{curtarolo:art144}.\par

The \AFLOW-\CCE\ workflow includes a structural analysis to identify
the number of cation-anion bonds, an automatic determination of
oxidation numbers based on Allen electronegativities, and the
inclusion of temperature effects by parametrizing vibrational
contributions to the formation enthalpy per bond.

With all the required functionality in place, the implementation will be extended to other anion classes beyond oxides such as nitrides, halides, and sulfides by adding the needed corrections in the near future.

\section{Code Availability} \label{code_avail}

\noindent
The Automated \CCE\ module is integrated into the \AFLOW\ software
(version 3.2.7 and later).
The source code is available at
\href{http://aflow.org/install-aflow/}{http://aflow.org/install-aflow/}
and
\href{http://materials.duke.edu/AFLOW/}{http://materials.duke.edu/AFLOW/},
and it is compatible with most Linux, macOS, and Microsoft operating systems.
The \CCE\ web tool is accessible via:
\href{http://aflow.org/aflow-online/}{http://aflow.org/aflow-online/}.
Tutorials are available through the \AFLOW-School:
\href{http://aflow.org/aflow-school/}{http://aflow.org/aflow-school/}.
Questions and bug reports should be emailed to \texttt{aflow@groups.io}
with a subject line containing ``\CCE''.

\section*{Acknowledgments}
\noindent
We thank Arkady Krasheninnikov, Demet Usanmaz, Frisco Rose,
Eric Gossett, Denise Ford, Andriy Smolyanyuk, and
Xiomara Campilongo for fruitful discussions.
The authors acknowledge support by DOD-ONR (N00014-16-1-2326, N00014-17-1-2090, N00014-17-1-2876), and by the National Science Foundation under DMREF Grant No. DMR-1921909.
R.F. acknowledges support from the Alexander von Humboldt foundation under the Feodor Lynen research fellowship.

\newcommand{\Ozolins}{Ozoli{\c{n}}{\v{s}}}

\end{document}